# A superhigh-frequency optoelectromechanical system based on a slotted photonic crystal cavity

Xiankai Sun, Xufeng Zhang, Menno Poot, Chi Xiong, and Hong X. Tang[a)]

*Department of Electrical Engineering, Yale University, 15 Prospect St., New Haven, Connecticut 06511, USA*

**Abstract:** We develop an all-integrated optoelectromechanical system that operates in the superhigh frequency band. This system is based on an ultrahigh-$Q$ slotted photonic crystal (PhC) nanocavity formed by two PhC membranes, one of which is patterned with electrode and capacitively driven. The strong simultaneous electromechanical and optomechanical interactions yield efficient electrical excitation and sensitive optical transduction of the bulk acoustic modes of the PhC membrane. These modes are identified up to a frequency of 4.20 GHz, with their mechanical $Q$ factors ranging from 240 to 1,730. Directly linking signals in microwave and optical domains, such optoelectromechanical systems will find applications in microwave photonics in addition to those that utilize the electromechanical and optomechanical interactions separately.

[a)]Electronic mail: hong.tang@yale.edu.

Nanoelectromechanical systems and nanooptomechanical systems that work at high frequencies have recently attracted great interest because of their use in studying mesoscopic quantum mechanics[1] as well as many practical applications such as high-speed sensing and coherent signal processing.[2-9] By combining these two types of systems into a single device where the mechanical degree of freedom is simultaneously coupled to both electrical and optical degrees of freedom, one achieves electro-optic interaction with capabilities that take advantage of the amplification from mechanical resonances. To this end, it is important to design a system possessing both strong electromechanical and optomechanical coupling such that the mechanical modes can be electrically excited efficiently and optically read out sensitively.

Thin-film bulk acoustic resonators (TFBAR) are one type of electromechanical systems that have been widely used in wireless applications, where the microwave-frequency bulk acoustic modes are usually excited by a piezoelectric element which converts electrical energy directly into mechanical energy.[10] This work pursues an implementation on an all-silicon platform because of its maturity in electronics and photonics integration. Without the piezoelectric effect in silicon, excitation of the bulk acoustic modes is implemented through the capacitive force between a pair of patterned metal wires, one of which locates on an edge of the free-moving mechanical resonator and the other is fixed. The electromechanical coupling is thus provided via the strong dependence of the capacitance on the nanosized gap between the metal-wire electrodes.[8] For optomechanical coupling, we employ slotted photonic crystal (PhC) nanocavities because of their simultaneous high optical quality ($Q$) factor and large optomechanical coupling.[11-16] The cavity is formed by a mode-gap confinement mechanism based on a variation from the corresponding slotted PhC waveguide. Among all the design schemes two are mostly implemented: one is tapering the lattice constant in the longitudinal direction[11-13] and the other is shifting the holes in the vicinity of the cavity center in the transverse direction.[14-16] We choose the latter design because of the convenience of incorporating access waveguides in the host PhC structure. Here, by combining an electrocapacitively actuated TFBAR with a slotted PhC nanocavity we develop an all-integrated optoelectromechanical system where the PhC membrane's bulk acoustic modes up to the 6th order at 4.20 GHz are efficiently excited and optically transduced. Apart from extending the operating frequency to the superhigh frequency band in many applications such as mechanical sensors and oscillators, our devices serve as a medium to directly link electromagnetic waves in microwave and optical domains and therefore has great potential in coherent signal processing.

The devices are fabricated from standard silicon-on-insulator (SOI) substrates (220-nm silicon layer on 3-μm buried oxide) with three major steps of lithography and subsequent processes. First, an electron-beam lithography lift-off process is used to make a patterned metal stack of Cr/Au (thickness 8 nm/80 nm) serving as the electrodes and alignment markers for the subsequent lithography steps. Second, the PhC structure with the routing strip waveguides is defined by a second electron-beam lithography step in ZEP 520A and then transferred to the silicon device layer with chlorine-based plasma dry etching. Third, by photolithography and a subsequent wet etch in a buffered oxide etchant, the oxide underneath the PhC structure is removed such that the PhC membranes are released from the substrate. The scanning electron microscope (SEM) image in Fig. 1(a) provides an overview of a fabricated device, including the electrodes, connecting strip waveguides, three membranes (labeled as "upper," "middle," and "lower"). The slotted PhC cavity comprises the middle and lower membranes, while the electrodes are patterned across the upper and middle membranes. This way excitation and transduction of the in-plane acoustic modes of the middle membrane can be achieved. Note that the SOI layer is compressively stressed, leading to slight buckling when the membranes are released.[17] However, with a similar mechanical stiffness, these membranes buckle in the same way after releasing thereby retaining the efficient drive and readout capability of the middle membrane. Figure 1(b) zooms in at the driving electrodes and capacitor slot, with designed widths of 600 nm and 150 nm respectively. Figure 1(c) shows the slotted PhC cavity consisting of the middle and lower membranes separated by an 80-nm cavity slot. The middle membrane is patterned with a metallic nanowire electrode for electrostatic actuation, while the lower membrane incorporates two PhC access waveguides in parallel with the cavity slot for optical input and output coupling. The color-coded holes are shifted laterally from their original lattice locations to form a mode-gap confined cavity, the design of which will be detailed below.

Figure 1(a) is a schematic of the experimental setup used for the device characterization. The chip is placed inside a vacuum chamber, which is pumped below $10^{-4}$ mbar to minimize the effect of the viscous gas damping on the mechanical resonator. Light from a C-band tunable diode laser is attenuated and polarization-adjusted before sent to the device. The light enters the chip via a grating coupler (not shown) and is routed toward the slotted PhC cavity. Light passing through the cavity is collected into an optical fiber via a second grating coupler. After a 99/1 fiber coupler, 1% of the transmitted light goes to a low-noise photodetector for monitoring the transmission level and for recording the transmission spectrum, while the other 99% is sent through a fiber preamplifier and a tunable filter before reaching a GHz photodetector. An AC driving voltage ($V_{ac}$) from a network analyzer is combined with a DC bias $V_{dc}$ using a bias-tee, and then applied onto the device electrodes to electrically control and actuate the middle PhC membrane that is a part of the slotted PhC cavity. Due to the large optomechanical coupling, the driven mechanical response of this PhC membrane is imprinted onto the transmitted optical signal, which is converted into electrical domain by the GHz photodetector and then sent back to the network analyzer for readout.

The slotted PhC cavity is formed by locally modulating the width of the corresponding slotted PhC waveguide.[14] Figure 2(a) plots the calculated TE-like band diagram of a slotted W1.2 waveguide (i.e., a PhC waveguide formed in the Γ–K direction of a triangular lattice where the waveguide width is $1.2\sqrt{3}a$ with $a$ the lattice constant). The lattice constant $a$ = 470 nm, hole radius $r = 0.29a$, membrane thickness $t$ = 220 nm, and slot width $w_s$ = 80 nm. Inside the bandgap exist four bands, the modes of which are guided by the slot. The first air-guided band with the highest frequency (red thick line) is used here for designing the cavity with a resonant wavelength around 1550 nm. The cavity is created by locally adjusting the width of the slotted PhC waveguide, where the red-, green-, and blue-marked holes in Fig. 1(c) are shifted away from the cavity slot in the lateral direction by 10.0 nm, 6.7 nm, and 3.3 nm, respectively. Such a gentle increase of waveguide width results in a local shift of the bands to the lower frequencies and creates a potential well in the longitudinal direction. This potential well supports the cavity modes by a mode-gap confinement mechanism.[14] Also included in the lower PhC membrane are two W0.84 waveguides, in parallel with the cavity slot, used to access the cavity mode in an in-line coupling configuration. The designed cavity structure without the PhC access waveguides is simulated with MEEP by a three-dimensional finite-difference time-domain method.[18] The cavity mode has a resonant wavelength $\lambda_0$ = 1544.8 nm, a simulated intrinsic optical $Q \approx 10^6$, and an effective modal volume $V_o = 0.043\ (\lambda_0)^3$. The electric field component $E_y$ in Fig. 2(c) clearly shows the strong field localization inside the cavity slot. The optomechanical coupling coefficient ($g_{om}/2\pi$) due to the moved membrane boundary is calculated to be 148 GHz/nm for the in-plane bulk acoustic modes, which value is similar to those reported elsewhere.[12, 13, 16]

The optical tunability of the slotted PhC cavity is characterized by performing a transmission spectroscopy measurement. As shown in Fig. 2(b), under a constant pulling electrostatic force the middle PhC membrane is displaced toward the electrode. This structural deformation leads to an increase of the cavity slot width and a consequent blue-shift of the cavity resonance due to the strong field localization inside the cavity slot. Figure 2(d) records the cavity transmission spectra under a DC bias from 0 to 9 V. The measured cavity linewidth of 10.6 pm corresponds to a loaded optical $Q$ of $1.44 \times 10^5$, which in combination with the measured cavity on-peak transmission gives an intrinsic optical $Q$ of $2.0 \times 10^5$. This $Q$ value is among the highest experimental demonstrations of slotted PhC cavities,[11-13, 15, 16] although a factor of 5 away from the theoretical expectation due to the unavoidable fabrication imperfections. With a maximally applied voltage of 9 V, the cavity resonance shift amounts to 217 pm, i.e., about 20.5 cavity linewidths. The tuning range is limited by the maximally applicable voltage on the electrodes. Typically, above 10 V the devices are permanently damaged. Figure 2(e) plots the cavity wavelength shift as a function of the applied DC bias. As theoretically predicted, the wavelength shift ($\Delta\lambda$) follows a quadratic relation with the applied DC voltage ($V_{\mathrm{dc}}$). If a cavity DC electrooptic tunability coefficient is defined as $\partial(\Delta\lambda)/\partial(V_{\mathrm{dc}}^2)$, a quadratic fit provides a value of −2.7 pm/V$^2$.

Next, the dynamical response of the PhC membrane under an AC capacitive drive is characterized by measuring the $S_{21}$ transmission spectrum from the network analyzer. To maximize the signal, the DC bias voltage is set at 8 V and the laser wavelength is set at the maximal slope of the cavity resonance peak under this DC bias. The optical power collected in the fiber at the output of the device is ~1 μW. Figure 3(a) shows a broad-range spectrum of the measured $S_{21}$ magnitude. The six dominant peaks ranging from 0.506 to 4.196 GHz, with an average separation of 0.738 GHz, are identified as the thin-film bulk acoustic modes (1st to 6th order) of the PhC membrane. Because these modes vibrate in plane and predominantly in the *y* direction (perpendicular to the capacitor slot), they are efficiently excited under the present electrostatic driving configuration. Note that the modes of the upper and lower membrane do not show up: the upper membrane is not coupled to the cavity and the lower one does not have actuation electrodes; only the in-plane motion of the middle membrane can be excited and detected. Figure 3(b)–(g) show the magnitude and phase response of the zoomed-in spectra of each acoustic mode, along with the corresponding mechanical displacement profiles calculated from finite-element simulation. By taking advantage of the modal translational symmetry (along the *x* direction), only a portion of the middle membrane as indicated by the dashed rectangle in the inset of Fig. 3(a) is simulated and displayed. It is evident that these acoustic modes satisfy the relation $W_{\mathrm{PhC}} = m\lambda_{\mathrm{a}}^{(m)}/2$ where $W_{\mathrm{PhC}}$ is the width of the PhC membrane (4.44 μm), *m* the mode order number, and $\lambda_{\mathrm{a}}^{(m)}$ the acoustic wavelength of order *m*. An acoustic mode can be efficiently excited only when its half acoustic wavelengths $\lambda_{\mathrm{a}}^{(m)}/2$ is much longer than the width of the driving electrode. Therefore, the modes of the 7th and higher orders cannot be efficiently excited because their half wavelengths are too short compared to the electrode width (600 nm). By fitting the magnitude spectra to a Lorentzian lineshape, the frequencies and mechanical *Q* factors are obtained for each mode and are summarized in Table I. The mechanical *Q* value varies from mode to mode and has a broad distribution from 240 to 1,730. These values are much better than those obtained from other PhC membrane structures measured in a similar environment,[19] which is attributed to the in-plane vibration nature of the bulk acoustic modes and a lower clamping loss of our structure. The mechanical *Q* factors could be further enhanced by shielding the slotted

PhC cavity with a two-dimension phononic bandgap structure to suppress the acoustic radiation to the substrate.[20]

In conclusion, we have developed a superhigh-frequency optoelectromechanical system on a CMOS-compatible all-integrated silicon photonics platform. By taking advantage of the large electrostatic driving force and strong optomechanical interaction from an ultrahigh-$Q$ slotted PhC nanocavity, the bulk acoustic modes of the PhC membrane are efficiently excited and optically read out. These acoustic modes are identified up to the 6th order at a frequency of 4.20 GHz, with clamping-limited mechanical $Q$ factors ranging from 240 to 1,730. Such optoelectromechanical systems provide a direct link between electromagnetic signals in the optical and microwave domains and are expected to find their applications in coherent signal processing in addition to low-phase-noise optoacoustic oscillators[9] and quantum ground-state cooling.[21]

We acknowledge funding from DARPA/MTO ORCHID program through a grant from the Air Force Office of Scientific Research (AFOSR). Facilities used were supported by Yale Institute for Nanoscience and Quantum Engineering and NSF MRSEC DMR 1119826. H.X.T. acknowledges support from the National Science Foundation CAREER award and a Packard Fellowship in Science and Engineering. M.P. thanks the Netherlands Organization for Scientific Research (NWO)/Marie Curie Cofund Action for support via a Rubicon fellowship. The authors thank Michael Power and Dr. Michael Rooks for assistance in device fabrication.

Figure captions:

FIG. 1. (a) Experimental setup: TDL, tunable diode laser; VOA, variable optical attenuator; FPC, fiber polarization controller; PD, photodetector; DAC, data acquisition card; EDFA, erbium-doped fiber amplifier (pre-amp); OTF, optical tunable filter; NWA, network analyzer. The scanning electron microscope (SEM) image shows a tilted view of the device, including the electrodes, slotted photonic crystal (PhC) cavity, and routing strip waveguides. (b) Top-view SEM image zooming in at the driving electrodes (width 600 nm) and the capacitor slot (width 150 nm). (c) Top-view SEM image showing the slotted PhC cavity and the PhC access waveguides in an in-line coupling configuration. The red-, green-, and blue-marked holes are shifted away from the cavity slot in the lateral direction by 10.0 nm, 6.7 nm, and 3.3 nm, respectively. The scale bars in the SEM images are 5 μm, 1 μm, and 1 μm in (a)–(c), respectively.

FIG. 2. (a) Band diagram of a slotted W1.2 PhC waveguide, formed in a 220-nm-thick silicon layer with lattice constant $a$ = 470 nm, lattice hole radius $r$ = 0.29$a$, and waveguide slot width $w_s$ = 80 nm. The continuum of PhC slab modes is indicated by the gray areas. Inside the PhC bandgap exist four guided waveguide bands, among which the highest one (red thick line) is used to form the cavity by a local modulation of the waveguide width [see Fig. 1(c)]. (b) Mechanical displacement profile of the slotted PhC cavity in response to an applied DC bias on the middle membrane. (c) Simulated TE-like electric field component $E_y$ of the resonant mode ($\lambda_0$ = 1544.8 nm) of the slotted PhC cavity. The zoom-in shows strong field confinement inside the slot. (d)

Optical transmission spectra of the slotted PhC cavity under various applied DC bias ($V_{dc}$) on the electrodes. The resonance linewidth corresponds to a loaded optical $Q$ of $1.44 \times 10^5$, or an intrinsic $Q$ of $2.0 \times 10^5$. (e) Cavity resonant wavelength shift as a function of the applied DC bias. The quadratic fit provides a cavity DC electrooptic tunability coefficient of $-2.7$ pm/V$^2$.

FIG. 3. (a) Magnitude of the $S_{21}$ transmission spectrum from the network analyzer, showing the excited bulk acoustic modes of the middle PhC membrane up to the 6th order at 4.20 GHz. The signal below 400 MHz is suppressed by a high-pass electrical filter to avoid overloading the network analyzer due to the large signal originating from the low-frequency mechanical modes of the membrane. (b)–(g) Zoom-in of the driven response of each mode showing the magnitude and phase. Insets: displacement profile of the corresponding acoustic mode (only the dashed rectangle region in the inset of (a) is shown due to the modal translational symmetry). Their frequencies and mechanical $Q$ factors are obtained from Lorentzian fitting (red) and are summarized in Table I.

TABLE I. Properties of the observed acoustic modes of the photonic crystal membrane.

| Mode No. | 1 | 2 | 3 | 4 | 5 | 6 |
|---|---|---|---|---|---|---|
| Measured frequency (GHz) | 0.506 | 1.254 | 1.971 | 2.720 | 3.460 | 4.196 |
| Simulated frequency (GHz) | 0.684 | 1.368 | 2.053 | 2.737 | 3.417 | 4.089 |
| Measured mechanical $Q$ | 470 | 240 | 1730 | 1420 | 650 | 240 |

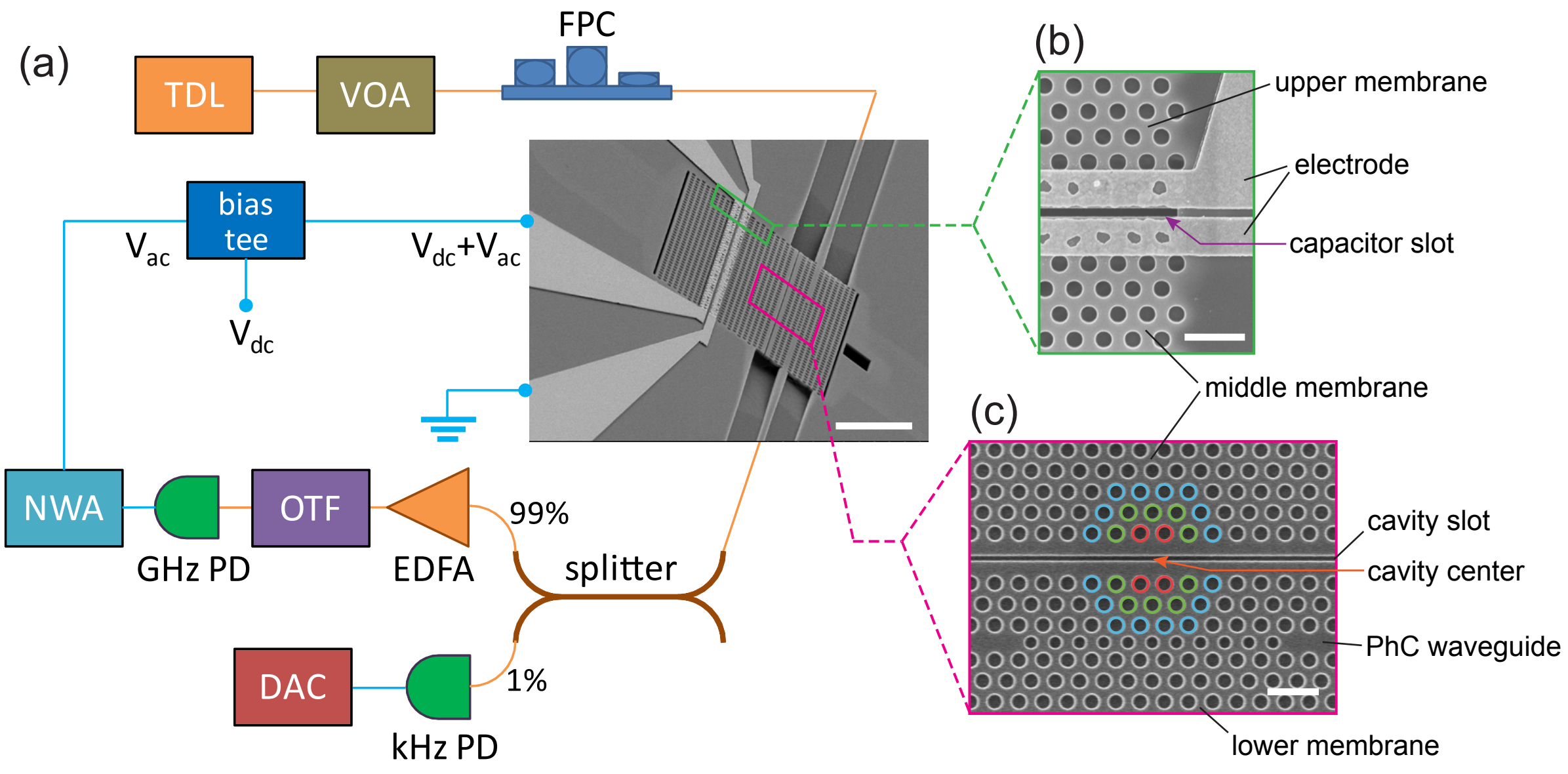

Fig. 1

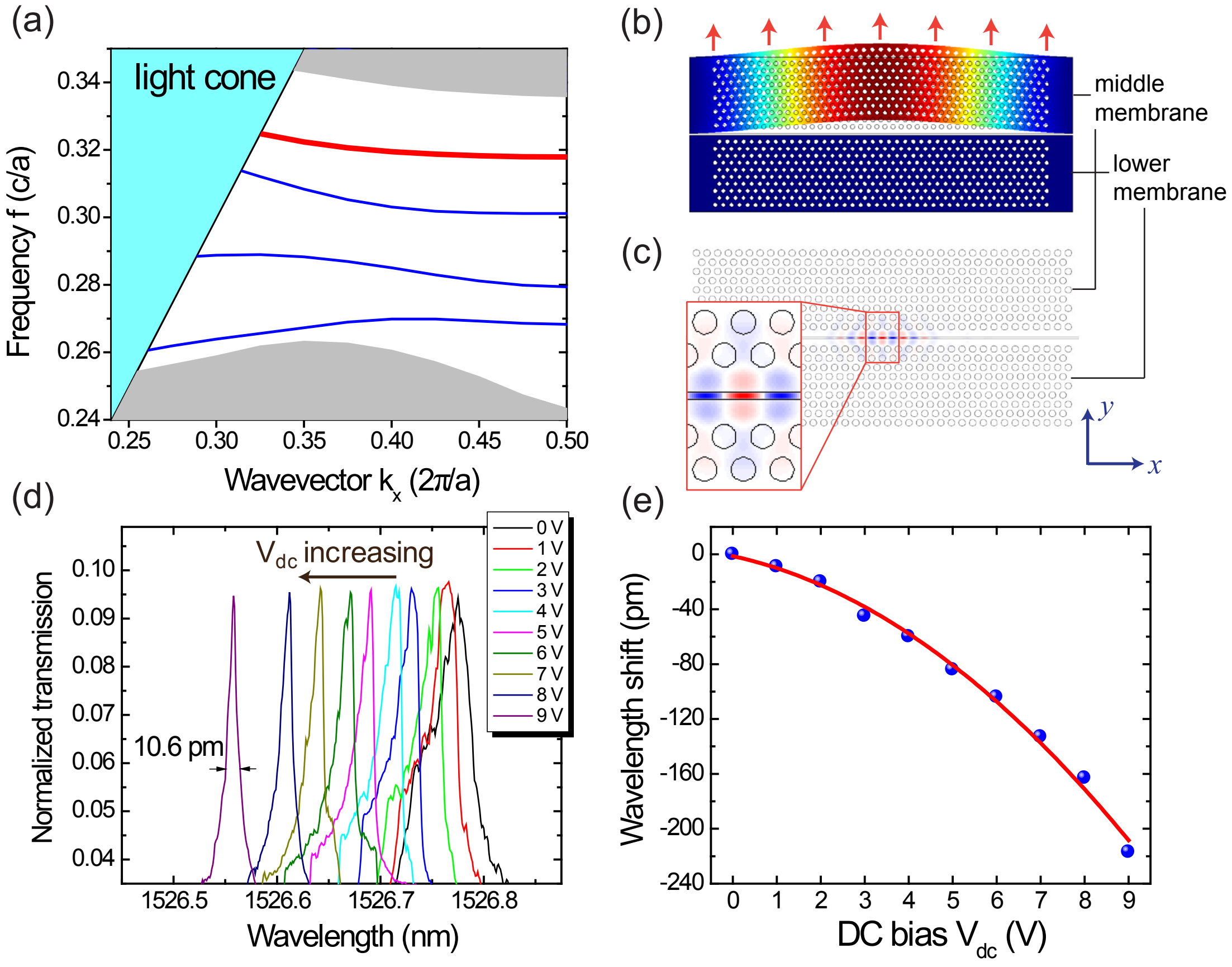

Fig. 2

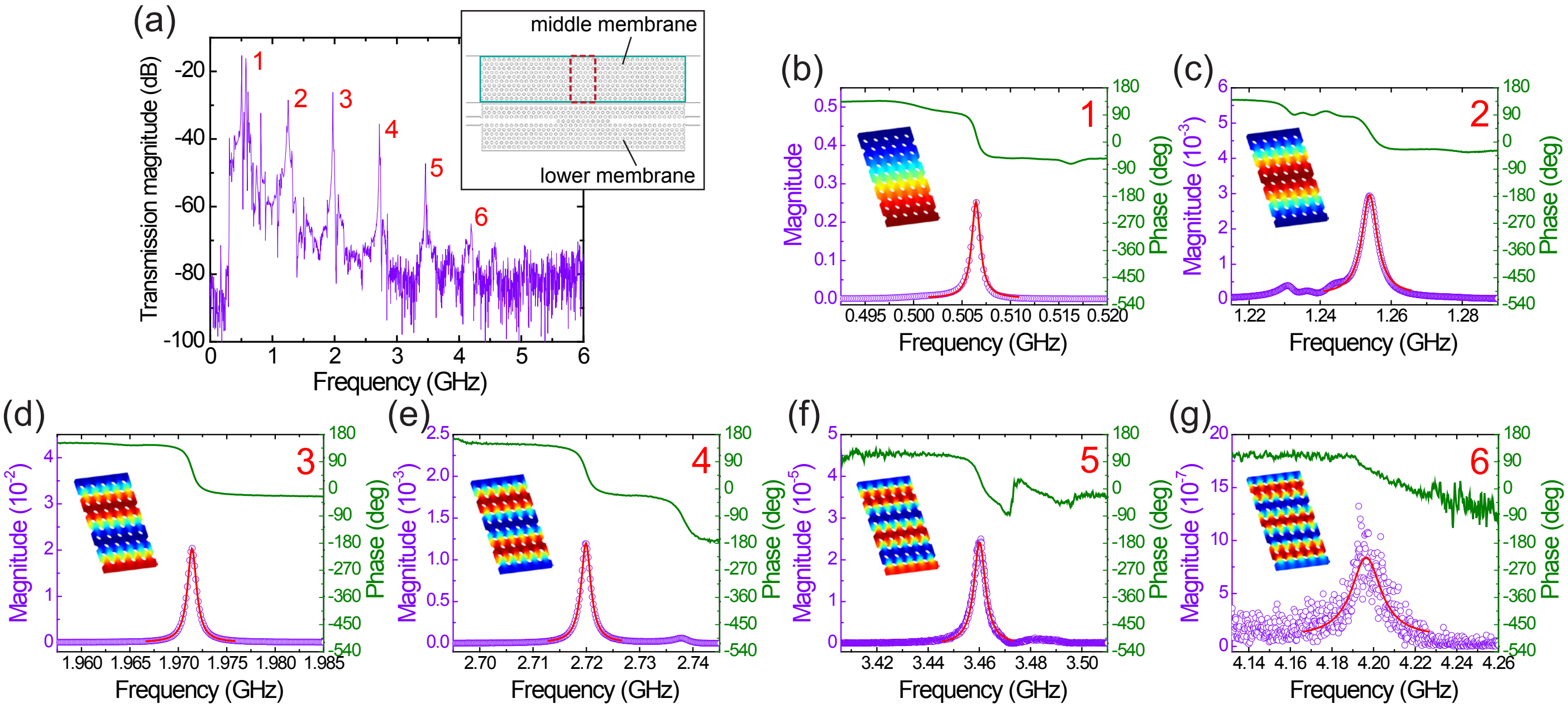

Fig. 3